\documentstyle[aps,prl,multicol,amssymb]{revtex}
\begin{document}
\draft
\preprint{}
\title{Magnetism and Transport for Two-Dimensional Electrons :
Absence of Stoner Ferromagnetism and Positive In-Plane Magnetoresistance
in the Metallic Phase}

\author{Philippe Jacquod}

\address {Instituut-Lorentz, Universiteit Leiden
P.O. Box 9506, 2300 RA Leiden, The Netherlands}

\date{\today}

\maketitle
\begin{abstract}
We calculate the interaction kernel $K$
for two-dimensional diffusive electrons. 
The screening of the Coulomb interaction together with the Fermi statistics
induces a spin selection rule for electron-electron scattering 
so that in leading order in the inverse conductance only pairs of 
electrons with antiparallel spins do scatter. 
At low temperature, this results in a larger
coherence length for fully polarized electrons and thus in a positive 
in-plane magnetoresistance. An applied in-plane
magnetic field also induces a nonmonotonous behavior of $K$ at
finite temperature. 
Alternatively, the vanishing of the scattering in the triplet channel
strongly reduces ferromagnetism deep in the metallic regime. 
These effects weaken as the density of charge carriers is reduced. 
\end{abstract}
\vspace{-2mm}

\vspace{-5mm}
\begin{multicols}{2}

Over the past five years, many experiments have reported
a nontrivial connection between transport
and magnetic properties of dilute, high-mobility two-dimensional
electron gases (2DEG). Especially intriguing is the
response of those systems to the application of a magnetic field
parallel to the confinement plane of the charge carriers.
The magnetoresistance in response to such a field
is enormous, leading to 
the suppression of the metallic behavior and the 
saturation of the resistance above a 
field strength $B_{\rm sat}$ \cite{simonian}. Shubnikov-de Haas
measurements indicate that this saturation occurs as
the electrons become fully spin polarized \cite{shdehaas}, while
no response was observed for systems with easy-spin
axis \cite{sige} : the effect is related to the electron spin.
Recent experiments report a significant increase of 
the magnetic 
susceptibility as the conductance of the sample
decreases \cite{gfactor} and it has even been reported
that $B_{\rm sat}$ vanishes as the conductivity crossovers from a metallic
to an insulating behavior where, extrapolated to $T=0$, an 
arbitrarily weak in-plane field seems sufficient to fully
polarize the 2DEG \cite{vitkalov} . Quite naturally, this raises the 
issue of the presence of a ferromagnetic instability
as the system becomes localized.

Strong connections between magnetic ordering and localization have been
known since the early days of quantum mechanics: 
Pairs of weakly overlapping, localized electrons can
reduce their energy via virtual (kinetic) exchange 
processes, which together with the Pauli exclusion,
leads to an effective antiferromagnetic coupling \cite{heitler}. 
For strongly overlapping,
itinerant electrons on the other hand, the interaction energy 
can be minimized 
by aligning the spins and this leads to an effective
ferromagnetic coupling \cite{heisenberg}. 
A Hartree-Fock treatment predicts then a ferromagnetic instability 
when the gain in exchange energy counterbalances the loss
in kinetic energy due to the promotion of electrons 
to higher unoccupied energy levels which leads to the {\it Stoner
criterion} \cite{stoner}

\begin{equation}\label{stonercrit}
J \equiv \int d^dr_{1} d^dr_{2} V_{1,2} \langle
\psi_{\alpha}^*(1) \psi_{\beta}(1)  
\psi_{\alpha}(2) \psi_{\beta}^*(2) \rangle = \Delta 
\end{equation}

\noindent Here $V_{1,2}=V(r_{1,2}\equiv|\vec{r}_1-
\vec{r}_2|)$ is the 
interaction potential, $\psi_{\alpha}(1) \equiv \psi_{\alpha}(\vec{r}_1)$
denotes a one-particle wavefunction and the average $\langle \ldots \rangle$
is taken over wavefunctions close to the Fermi level. $\Delta$ is the
level spacing at the Fermi level. 

The threshold (\ref{stonercrit}) overestimates the tendency 
toward magnetic ordering : it
is a perturbation result {\it in the interaction}
which often predicts a threshold where electrons repel each other so strongly 
that a perturbation theory {\it in the kinetic energy} is more adequate.
In most cases the ferromagnetic order is then destroyed by antiferromagnetic 
superexchange \cite{anderson,pichard}.
In other instances, 
$J < \Delta$ and paramagnetism prevails. Remarkable exceptions
are provided by materials with orbital degeneracies for which ferromagnetism
is favored by the absence of any kinetic energy cost for spin alignment.
This is the essence of Hund's first rule and is also believed to
lead to ferromagnetism in Fe, Co and Ni \cite{vanvleck}. 

Roughly speaking, itinerant 
electrons have a tendency toward ferromagnetism, while the coupling
becomes antiferromagnetic between localized electrons or in the case
of a strong local repulsion. Below we will
establish a reversed connection between magnetic order and 
localization for disordered systems 
where ferromagnetism is suppressed beyond the Thouless energy 
$E_{\rm c}=g \Delta$ ($g$ is the conductance) in the
delocalized phase while the mechanism behind this suppression
weakens as the system becomes insulating, resulting in an increasing
magnetic susceptibility in qualitative agreement with 
experiments \cite{gfactor,vitkalov} and numerical simulations \cite{dima}.

For diffusive metals, extending the
Hartree-Fock treatment of the Stoner ferromagnetism 
seems appropriate : the (kinetic)
antiferromagnetic superexchange
is damped by (nonmagnetic) impurity scattering.
Simultaneously, the presence of disorder induces
stronger wavefunction correlations, effectively
increasing the strength of the ferromagnetic exchange (\ref{stonercrit})
whereas it strongly reduces by a factor
$O(1/g) \ll 1$ the typical magnitude of 
off-diagonal interaction matrix elements
\cite{altshuler,mirlin}. It thus seems reasonable to truncate
the perturbation expansion in the interaction. Doing so at first
order predicts an increased probability for 
weak but finite ground-state spin polarization due to 
disorder \cite{andreev}, a 
conclusion that has found experimental confirmation in
ballistic microcavities \cite{dotexp}.
When one considers higher order corrections in the interaction
one finds however that in either the clean \cite{perthub} or
in the disordered 
case \cite{stone}, Hartree-Fock approaches overestimate 
the magnetic ordering. In this letter we will amplify on these latter results
by a microscopic calculation of the relevant parameters of the 
perturbative expansion developed in \cite{stone}. 
We will see that in a diffusive
metal, and assuming a Spin Rotational Symmetric (SRS) Coulomb 
interaction, {\it electrons with parallel spins do not scatter} in leading
order in $1/g$, which
leads to a strong demagnetizing effect in the metallic
diffusive regime. 

Our arguments can be summarized as follows.
SRS implies that interaction-induced transitions occur between pairs of
fermions with conserved total spin ($j=0$ or 1).
Fermi statistics requires the 
symmetrization (antisymmetrization) of the off-diagonal matrix 
elements connecting $j=0$ ($j=1$) two-particle states. For electrons 
with parallel spins, this prescription results in the vanishing of the
interaction for distances shorter than the Fermi
wavelength $r \le k_F^{-1}$ and this strongly reduces off-diagonal
interactions for triplet-paired electrons. The effect becomes
stronger as $k_F$ is reduced: writing the inverse screening length as 
$\kappa_2 \equiv \alpha_2 k_F$, with
$\alpha_2=1/(k_F a) >1$ ($a$ is the Bohr radius), 
one has, ($g_v$ is the
valley degeneracy), $\alpha_2 = \sqrt{2} g_v^{3/2} r_s$ \cite{stern}. 
$r_s$ is the gas 
parameter, hence $\alpha_2$ increases as $k_F$ is reduced.

Three important consequences 
are : (i) Extending the second order perturbation
theory developed in \cite{stone} we get a much larger 
energy cost for polarizing spins and this 
suppresses the Stoner instability beyond the Thouless energy.
At lower $g$, this
energy cost is reduced and the magnetic susceptibility increases
as $g \rightarrow 1$. This effect is enhanced by a 
simultaneous increase of the ferromagnetic coupling constant 
so that extrapolating our
results to $g \approx1$ (where however the theory loses its validity) 
we reach a ferromagnetic instability. 
(ii) The strongly reduced scattering for electrons with parallel 
spins increases
the coherence length at low temperature for polarized electrons.
As an in-plane magnetic field is applied to a 2DEG, this is 
accompanied by an increase of weak localization corrections
and thus by positive magnetoresistance. (iii)
In presence of a magnetic field, the Zeeman splitting $\Delta \omega = B$
results in an energy shift for the interaction kernel which becomes:
$K(\omega;\uparrow/\downarrow) = K(|\omega \pm B|)$, 
where the ``+'' and ``-''
signs correspond to an electron with excitation energy $\omega$
being in the majority ($\uparrow$) and minority ($\downarrow$) 
spin subband respectively. Accordingly, the average
$\tilde{K}=(K(\omega;\uparrow)+K(\omega;\downarrow))/2$ 
shows a nonmonotonous behavior at finite excitation energy (or
temperature) $\omega$, 
increasing with $B$ for $B<\omega$ and decreasing again for $B>\omega$.

We now present our calculations.
Under requirement of SRS and rotational symmetry in Hilbert space   
a generic Hamiltonian for interacting electrons in a weakly disordered 
metallic system can be written \cite{stone,kurland}

\begin{equation}
    {\cal H} = {\cal H}_0 - J \vec{S} \cdot \vec{S}
    +{\cal U}_{f}.
\end{equation}

\noindent The first term is a one-body, spin independent Hamiltonian with
eigenfunctions $\psi_{\alpha}$.
The second term represents the ferromagnetic exchange,
$J >0$. We have introduced spin operators
$\vec{S}_{\alpha} \equiv (1/2)\sum_{s,t}
c^{\dagger}_{\alpha,s} \vec{\sigma}_{s,t} c_{\alpha,t}$ and
$\vec{S} = \sum_{\alpha} \vec{S}_{\alpha}$, where the $c$'s are 
fermionic operators. The third term finally 
contains the physics beyond the mean-field approach,
i.e. the fluctuating part of the interaction

\begin{eqnarray}
{\cal U}_{f} & = & \sum_{\alpha,\beta;\gamma,\delta}\sum_{s,s'}
    U_{\alpha,\beta}^{\gamma,\delta}
c^{\dagger}_{\alpha,s} c^{\dagger}_{\beta,s'} c_{\delta,s'}
c_{\gamma,s}, 
\end{eqnarray}

\noindent where the interaction matrix elements are defined as

\begin{eqnarray}\label{ime}
U_{\alpha,\beta}^{\gamma,\delta} & = & \int d^{2}r_{1} d^{2}r_{2}
V_{1,2} \psi_{\alpha}^*(1)  \psi_{\beta}^*(2)
      \psi_{\delta}(2) \psi_{\gamma}(1).
\end{eqnarray}

\noindent We will use the $d=2$ screened Coulomb potential 

\begin{eqnarray}\label{rpacoulomb}
    V_{1,2}= (e/2 \pi)^2 \int d^2 q \exp(i \vec{q} \cdot 
\vec{r}_{1,2})/(q+\tilde{\kappa}_2(q,\omega)),
\end{eqnarray}

\noindent where screening depends on the energy transfer $\omega$,
$\tilde{\kappa}_2(q,\omega=0) = \kappa_2$ \cite{altshuler} and we
wrote $\vec{r}_{1,2}=\vec{r}_{1}-\vec{r}_{2}$.

${\cal U}_{f}$ can be rewritten as a sum over 
four contributions : the first one inducing scattering between 
singlet-paired ($j=0$) and the other three between
triplet-paired ($j=1$, $M=0,\pm 1$) two-electron states,
defining electron-electron scattering in {\it singlet and triplet channels}
respectively.
Accordingly, the interaction matrix elements must be symmetrized or 
antisymmetrized corresponding to the singlet (``$+$'' sign in the 
formula below) and the triplet (``$-$'' sign) channel
respectively

\begin{eqnarray}\label{symasym}
U_{\alpha,\beta;\gamma,\delta}^{\pm} & = & U_{\alpha,\beta}^{\gamma,\delta}+
U_{\beta,\alpha}^{\delta,\gamma} \pm U_{\alpha,\beta}^{\delta,\gamma} 
\pm U_{\beta,\alpha}^{\gamma,\delta}.
\end{eqnarray}

\noindent From (\ref{ime}) and (\ref{symasym}), the interaction 
matrix elements
in both channels are symmetrically distributed around zero average
and the interaction kernel is related to the variance
of these distributions $K_{\pm} \equiv \sigma^2(U^{\pm})$.
${\cal U}_{f}$ systematically opposes ferromagnetism, as it
induces more virtual processes in low-spin 
sectors \cite{stone}.
Since in the ferromagnetic sector perturbative corrections to the energy 
contain only contributions from the triplet channel, this tendency is further 
amplified
if $K_- \ll K_+$. We will now show that this is indeed the case
in a metal.

After disorder averaging, $K_+$
and $K_-$ are given by
\end{multicols}

\begin{eqnarray}\label{long}
K_{\pm} = 8 \langle U_{\alpha,\beta}^{\gamma,\delta} (
(U_{\alpha,\beta}^{\gamma,\delta})^{*} \pm 
(U_{\beta,\alpha}^{\gamma,\delta})^{*}\rangle & = &
8 \int \prod_{i=1}^{4} d^{d} r_{i} V_{1,2} V_{3,4} 
\left[
\langle \psi_{\alpha}^*(1) \psi_{\gamma}(1) 
\psi_{\alpha}(4) \psi_{\gamma}^*(4) \rangle 
\langle \psi_{\beta}^*(2) \psi_{\delta}(2) 
\psi_{\beta}(3) \psi_{\delta}^*(3) \rangle \right. \nonumber \\
& & \left.
\pm 
\langle \psi_{\alpha}^*(1) \psi_{\gamma}(1) 
\psi_{\alpha}(3) \psi_{\gamma}^*(4) \rangle 
\langle \psi_{\beta}^*(2) \psi_{\delta}(2) 
\psi_{\beta}(4) \psi_{\delta}^*(3) \rangle
\right].
\end{eqnarray}
\vspace{-0.7cm}
\begin{multicols}{2}

\noindent For diffusive systems, $l_e \ll L$ 
($l_e$ is the elastic mean free path, $L$ the linear system
size), one has up to $O(1/g)$ \cite{mirlin} 

\begin{eqnarray}\label{wavefn}
& & \langle
\psi_{\alpha}^*(1) \psi_{\gamma}(1) 
\psi_{\alpha}(4) \psi_{\gamma}^*(4) \rangle = \Pi(r_{1,4})/\Omega^2.
\end{eqnarray}

\noindent $\Omega=L^d$ and the diffusion propagator is given by 
$\Pi(r) = 1/(4 \pi^3 \nu) \int d^2q \exp(i \vec{q} \cdot 
\vec{r})/(Dq^2-i \omega)$.
$D=g/(2 \pi \nu)$ is the classical diffusion constant, $\nu$ is the density
of states, the integral ranges from $L^{-1}$ to $l_e^{-1}$
and $\omega$ gives the
energy difference between the states $\alpha$ and $\gamma$.
When $r_{3,4} < k_F^{-1}$ the second term between brackets in the
integrand of (\ref{long}) is equal to the first one and we can readily
conclude that for a strong screening $\alpha_2 \gg 1$
the triplet channel scattering is completely suppressed (this limit
corresponds to the Hubbard model).
For $r_{3,4} > k_F^{-1}$, one has 

\begin{eqnarray}\label{ballistic}
\langle \psi_{\alpha}^*(1) \psi_{\gamma}(1) 
\psi_{\alpha}(3) \psi_{\gamma}^*(4) \rangle \propto
\exp(-(r_{1,3}+r_{1,4})/(2 l_e)),
\end{eqnarray}
 
\noindent and (\ref{ballistic}) vanishes
exponentially for $l_e \ll L$. We write
$K_\pm = K \pm \delta K$. A straightforward calculation
gives
$K = 2 (\Delta/\pi^2 g)^2$. To calculate $\delta K$ we note
that, in the strongly diffusive regime the diffusion modes
are strongly damped for $q > l_e^{-1}$, furthermore
one has $\kappa_2 l_e \gtrsim k_F l_e \gg 1$. This justifies an 
expansion of the
Coulomb potential (\ref{rpacoulomb}) in inverse powers of $\kappa_2$.
Keeping only the first two nonvanishing orders (which is consistent with 
the above expression for the diffusion
propagator and the truncation of (\ref{wavefn}) at the first order 
in $1/g$) one gets for $\omega < E_{\rm c}$

\begin{eqnarray}\label{sigma}
\delta K & = & K (1-4(\alpha_2 g)^{-1} (l_e/L))+O(g^{-4}),
\end{eqnarray}
 
\noindent and for $\omega > E_{\rm c}$, both
$K$ and $\delta K$ must be multiplied
by a factor $(1+ \omega/E_{\rm c})^{-1}$ \cite{altshuler}.
$K_-$ is thus suppressed by a factor $(\alpha_2 g)^{-1} l_e/L \ll 1$
compared to $K_+$ which reflects the fact that the long-range
part of the RPA potential (\ref{rpacoulomb}) gives subdominant 
contributions to $K$ as long as $\kappa_2 l_e \gg 1$, and we 
thus set $K_-=0$. Note that for $d=3$,
the discrepancy between $K_+$ and $K_-$ is amplified
by the exponential screening of the interaction while ballistic systems have
$K_+/K_- = O(1)$ \cite{jacquod}.

Up to $O(1/g)$, the average exchange strength (\ref{stonercrit}) reads

\begin{eqnarray}\label{jexch}
J = (\Delta/2) [\tilde{J}_{SR} + \tilde{J}_B + \tilde{J}_D]
\end{eqnarray}

\noindent $\tilde{J}_{SR}$, $\tilde{J}_B$ and $\tilde{J}_D$ are 
short-range (for
$r_{1,2}<1/\kappa_2$ in (\ref{stonercrit})), long range
($r_{1,2}>1/\kappa_2$) ballistic 
and diffusive contributions 
respectively 

\begin{eqnarray}\label{JS}
\tilde{J}_{SR} & = & 1+\log(L/l_e)/(\pi g) \nonumber \\
\tilde{J}_B & = & \int_{\alpha_2^{-1}}^{k_F l_e} dR
J_0^2(R)/(\alpha_2 R^2)\\
\tilde{J}_D & = & (1+\tilde{J}_B) \log(L/l_e)/(\pi g)  \nonumber 
\end{eqnarray}

\noindent For $k_F l_e = g \gg 1$,
$\tilde{J}_B(\alpha_2) \approx 
1-(1-\pi^{-1})/\alpha_2$. Interestingly for $\alpha_2 \gg 1$, 
one has $J > \Delta$
and a first order treatment predicts a ferromagnetic (Stoner) instability.

We can now calculate the average
ground-state energy in each spin sector up to the second order in the
interaction. For fixed $S$, the number of majority (+) and minority (-) spins
are $n_{\pm}=n/2 \pm S$ where $n$ is the 
total number of electrons. We first neglect energy dependences
in (\ref{rpacoulomb}-\ref{wavefn}) 
which reduce $K_\pm$ and accordingly truncate the Hamiltonian
at the Thouless energy, i.e. 
we set $n = g$. The number of singlet (triplet) transitions in leading 
order is then $N_+=g^2 n_{+} 
n_{-}/4$ ($N_-=g^2 (n_{+}^2+n_{-}^2+n_{+} 
n_{-})/4$) where the prefactors $g^2$ account for the
final states. In presence of a 
magnetic field $B$, 
the average energy difference between the paramagnetic 
ground-state and the ground-state in the $S$-sector in second order
perturbation theory is given by

\begin{eqnarray}\label{2ndorder}
\Delta_S & = & J S(S+1) + B S -S^2 \left[\Delta 
+ g \log(g) \delta K/(2 \Delta) \right]
\end{eqnarray}
\vspace{-0.4cm}

\noindent which illustrates the
demagnetizing influence of $\delta K$. For large metallic samples
with $g \gg 1$ and $l_e \ll L$ one has $J \le \Delta$ and the magnetization
is bounded by $S_0 \le S_{\rm max} = O(g/\log(g))$. Electrons
are not polarized beyond the Thouless energy $E_c$ and
hence there is no thermodynamic
magnetization. 
At the same time this does not rule out the probability
to find few polarized electrons in agreement with experiments on
quantum dots in the Coulomb blockade regime \cite{dotexp}. 
(In that latter case however, ballistic contributions that were
neglected in (\ref{wavefn}) have to be considered \cite{jacquod}.)
At lower $g$ on the other hand, 
one may get $J>\Delta$ ($\alpha_2$ increases), and since generally 
$\log(L/l_e) > \log(g)$ for $d=2$, the polarization does not
saturate below $E_c$. For larger energies $\omega > E_c$, because
of the decay of $K$, second order corrections
become marginal, and therefore the polarization becomes thermodynamical.
While this latter point should be taken with a grain of salt
(even nonmagnetic materials may have an exchange above the Stoner 
threshold (\ref{stonercrit}),
see discussion in the introduction), this argument demonstrates that
the magnetic susceptibility increases as $g$ decreases.

We next calculate the magnetic field $B_{\rm sat}$
necessary to achieve full polarization ($\Delta B_{\rm sat}
\equiv B_{\rm sat}-B_{\rm sat}^0$; $B_{\rm sat}^0=2 g(\Delta-J)-J$ is 
the first-order saturating field). For $g \gg 1$, $\Delta B_{\rm sat}$
is dominantly given by the field it takes to polarize the 
electrons up to the Thouless energy:

\begin{eqnarray}
\Delta B_{\rm sat} & \approx & 
4 \Delta (1-4(\alpha_2 g)^{-1} (l_e/L)) \log(g)/\pi^4.
\end{eqnarray}

\noindent 
Clearly, $B_{\rm sat}$
decreases when $g$ is reduced in agreement with the experiments
\cite{gfactor,vitkalov}.

We finally calculate the electron-electron scattering rate for an 
electron at an excitation energy $\omega$ above a Fermi sea 
polarized by an external magnetic field $B$. For an electron in a given 
spin subband (the $+$ and $-$ signs correspond
to the majority and minority subband respectively)
this rate is determined by the scattering with
a particle below the Fermi energy in the other spin subband
with an energetic difference of $\omega \pm B$. Consequently, $K$
becomes spin-dependent,  
$K(\omega) \rightarrow K(\omega;\uparrow/\downarrow)=
K(0)(1+|\omega \pm B|/E_c)^{d/2-2} $
and for $d=2$ one gets an average scattering 
rate $\Gamma=(\Gamma_+ + \Gamma_-)/2$ with 

\begin{eqnarray}\label{gammae}
\Gamma_{\pm}(B,\omega) & = & (2 \pi/\hbar) 
K(\omega;\uparrow/\downarrow) \nu_3(\omega) \\
& = & (2 \pi/\hbar) K(0) \nu_3(\omega) (1+|\omega \pm B|/E_c)^{-1}
\end{eqnarray}

\noindent $\nu_3$ is the three-particle
density of states which we assume is spin-independent. For 
$B > E_c$, one then has a nonmonotonous behavior
of $\Gamma(B,\omega)$ vs. $\omega$
which in particular develops a local maximum at $\omega=B$ for $B \gtrsim 
2.28 E_c$. This is so because of the nonmonotonous behavior
of the average interaction kernel ${\tilde K}(\omega)=(K(\omega;\uparrow)
+K(\omega;\downarrow))/2$)

\begin{eqnarray}\label{nonmon}
\partial {\tilde K}(\omega)/\partial \omega  > 0 \ \ \ ; \ \ \ \omega < B,
\nonumber \\
\partial {\tilde K}(\omega)/\partial \omega  < 0 \ \ \ ; \ \ \ \omega > B.
\end{eqnarray}

\noindent At low temperature ($T=\omega$), 
dephasing is primarily due to electron-electron scattering, and 
for $\omega,B \gg E_c$
the weak-localization correction to the conductivity read

\end{multicols}

\begin{eqnarray}\label{weakloc}
\delta \sigma_B/\delta \sigma_0 & = & 1+
\{\log[\omega/2] +\log[1/(\omega + B)+1/|\omega - B|]\}/
\log[\Gamma(0,\omega) \tau_e],
\end{eqnarray}
\vspace{-0.7cm}
\begin{multicols}{2}

\noindent where $\tau_e$ is the elastic mean free time.
Equations (\ref{gammae} - \ref{weakloc}) predict
a nonmonotonous behavior of the quantum corrections to the
conductivity as a function of an in-plane magnetic field.
Accordingly, the conductivity first increases at low field until
$B = \omega$ and then decreases for larger fields. This should
dominate the behavior of the conductance for low enough temperature
and good metallic samples. 

Experimentally, the interaction kernel $K$
can be obtained via measurements of the energy distribution of 
quasiparticles as e.g. in metallic quantum wires \cite{devoret}.
The geometry
of those systems differs from the twodimensional case studied
here, resulting in a different energetic decay of $K(\omega)$
\cite{altshuler}. However as long as the magnetic field induces 
only a Zeeman coupling one expects 

\begin{equation}\label{nonmon2}
\tilde{K}(B,\omega) \propto (\omega+B)/E_c)^{a}+
(|\omega-B|/E_c)^{a}.
\end{equation}

\noindent Recently, such a behavior 
has been observed for copper wires \cite{anthore}. 
While the presented theory does not explain why experimentally
$a \ne -3/2$ as expected \cite{altshuler}, (\ref{nonmon2}) nicely
reproduce the experimentally obtained nonmonotonous behavior of 
$K(\omega)$ (see Fig. 5 in \cite{anthore}). Alternatively,
the nonmonotonicity of $K(\omega)$ could be
due to magnetic impurities (which would also explain the anomalous behavior 
$K(\omega) \propto (E_c/\omega)^{a}$ with 
$a \ne d/2-2$) \cite{anthore,glazman}, therefore
similar experiments on silver samples where $a= d/2-2$ and thus
magnetic impurities do not play a dominant role \cite{devoret} would
be highly desirable. A detailed calculation for the geometry corresponding
to those experiments will be published elsewhere \cite{jacquod}.

In summary we have found that spin selection
rules for electron-electron scattering in diffusive metals
induce a nontrivial coupling between transport and magnetic properties of
2DEG and a nonmonotonic behavior of the dephasing rate in presence
of a Zeeman coupling.

Work supported by the Swiss National Science Foundation and the Dutch
Science Foundation NWO/FOM. Interesting 
discussions with C. Beenakker, Ya. Blanter, P. Denteneer, 
F. Gebhard, J.-L. Pichard and J. Zaanen are gratefully acknowledged.

\end{multicols}
\end{document}